\documentclass{mem}
\usepackage{natbib}\usepackage{txfonts}\usepackage{balance}
\usepackage{graphicx}
\usepackage[a4paper,breaklinks,dvipdfm]{hyperref}
\idline{75}{282}
\begin{document}
\def\teff{$T\rm_{eff }$}
\def\kms{$\mathrm {km s}^{-1}$}
\def\Mprim{M_{1,\mathrm{i}}}
\def\Msec{M_{2,\mathrm{i}}}
\def\ai{a_{\mathrm{i}}}
\def\qi{q_{\mathrm{i}}}
\def\snia{SNe~Ia}
\def\msun{M_{\odot}}
\def\rsun{R_{\odot}}
\def\dtd{\mathrm{DTD}}
\def\SDH{SD$_{\mathrm{H}}$}
\def\SDHe{SD$_{\mathrm{He}}$}
\def\actaa{Acta Astron.}
\def\nar{New A Rev.}

\title{
What is the role of wind mass transfer in the progenitor evolution of Type Ia Supernovae?
}

   \subtitle{}

\author{
Carlo Abate
}

\institute{
Argelander-Institut f\"ur Astronomie -- Universit\"at Bonn,
Auf dem H\"ugel 71,
D-53121 Bonn, Germany
\email{cabate@uni-bonn.de}
}

\authorrunning{C.\,Abate }

\titlerunning{What is the role of wind mass transfer in the progenitor evolution of SNe Ia?}

\abstract{
Type Ia supernovae (SNe Ia) are thermonuclear explosions of carbon-oxygen white dwarfs (WDs) that accrete mass from a binary companion, 
which can be either a non-degenerate star (a main-sequence star or a giant) or an other WD in a binary merger 
(single- and double-degenerate scenario, respectively).  In population-synthesis studies of \snia\ the contribution of asymptotic 
giant branch (AGB) stars to either scenario is marginal. However, most of these studies adopt simplified assumptions to compute 
the effects of wind mass loss and accretion in binary systems. This work investigates the impact of wind mass transfer on a 
population of binary stars and discusses the role of AGB stars as progenitors of \snia. 
}
\maketitle{}

\section{Introduction}
\label{sec:intro}

It is generally accepted that type Ia supernovae (SNe Ia) are thermonuclear explosions of carbon-oxygen white dwarfs \cite[][]{Hoyle1960}. 
The explosion occurs when the WD reaches a critical mass close to the Chandrasekhar mass
($\approx\!\!1.4\,\msun$ for non-rotating stars) by transferring material from a binary companion.
Two main evolutionary paths have been proposed to describe this process.
In the \emph{single-degenerate (SD) scenario} the WD steadily accretes mass from a non-degenerate star, such as a main sequence star 
or a giant \cite[][]{Whelan1973, Nomoto1982}. In the \emph{double-degenerate (DD) scenario} two carbon-oxygen WDs merge and 
their combined mass exceeds the Chandrasekhar mass \cite[][]{Webbink1984, Iben1984}.
\cite{Hillebrandt2000} provide an exhaustive review of the two scenarios
and discuss the discrepancies between modelled and observed properties of \snia.

One problem of the models is that binary population-synthesis (BPS) studies typically predict rates of \snia\
explosions several times lower than the observations \cite[][]{Toonen2012,Claeys2014}. In these models, the contribution of 
asymptotic giant branch (AGB) stars is marginal, for two reasons. First, because Roche-lobe overflow (RLOF) from AGB 
donor stars is mostly unstable \cite[][]{Paczynski1965}, AGB stars do not efficiently transfer mass on to WDs 
in the SD channel. Second, in the DD scenario only AGB stars in a relatively narrow range of separations form WDs
in orbits sufficiently tight to merge within a Hubble time. Consequently, only about $10\%$ of
all possible systems producing a \snia\ event pass through (at least) one AGB phase \cite[][]{Claeys2014}.
However, most BPS models assume spherically-symmetric winds and low wind accretion efficiencies, as predicted by 
the canonical Bondi-Hoyle-Lyttleton (BHL) model. In contrast, recent hydrodynamical simulations of binary systems 
with AGB donors show that AGB winds can be very efficiently accreted by the companions in some circumstances
\cite[e.g. in the wind-RLOF, or WRLOF, regime described by][]{Shazrene2012}, and that the expelled 
material can carry a significant amount of angular momentum, in some cases shrinking the orbit 
significantly \cite[][]{Brookshaw1993}.

This study investigates the impact of the high wind-accretion efficiencies and the strong 
angular-momentum losses on the \snia\ rate.

\section{Models}
\label{sec:models}
The BPS code \texttt{binary\_c/nucsyn} \cite[][]{Izzard2004} is used to simulate the evolution
of binary systems for a wide range of initial primary and secondary masses ($\Mprim$ and $\Msec$, respectively)
and initial orbital separations ($\ai$). In each simulation the evolution of $150 \times 150 \times 200$
binary systems in a $\log \Mprim$--$\log \Msec$--$\log \ai$ parameter space is calculated.

The delay-time distribution (DTD), that is the \snia\ rate as a function of time per unit mass 
in stars formed in an initial starburst at $t=0$, is computed as 

	\begin{eqnarray}
		\dtd(t) = \frac{
		\sum\limits_{\Mprim}\sum\limits_{\Msec}\sum\limits_{\ai} \delta_{\mathrm{SNeIa}}~ \Psi_{M_1,M_2,a}~ \delta\!V}%
		{M_{\mathrm{total}}~\delta t}%
		, \label{eq:snia}
	\end{eqnarray}
%
where:
	\begin{itemize}
		\item $\Mprim$, $\Msec$ and $\ai$ vary in the intervals $[2.5,\!9.0]\msun$, $[1.0,\!\Mprim]\msun$, and $[5,\!10^5]\rsun$, respectively,
		\item $\delta_{\mathrm{SNeIa}}=1$ if the binary system ends its evolution with a \snia\ event within the 
				time interval $[t,~t+\delta t]$, and is zero otherwise.
		\item $\Psi_{M_1,M_2,a}$ is the initial distribution function of $\Mprim$, $\Msec$, $\ai$ and is assumed to be 
				separable. Consequently, $\Psi_{M_1,M_2,a}=\psi(\Mprim)\phi(\Msec)\chi(\ai)$, where $\psi(\Mprim)$ 
				is the solar-neighbourhood initial mass function of \cite{Kroupa1993}, $\phi(\Msec)$ is the initial distribution of 
				secondary masses, which is considered to be constant in $\qi=\Msec/\Mprim$, and $\chi(\ai)$ is the initial distribution of 
				orbital separation, which is assumed flat in $\log \ai$ \cite[][]{Opik1924}.
		\item $\delta\!V = \delta \Mprim~\delta \Msec~ \delta \ai$ is the volume of a cell in the parameter space of the grid.
		\item $M_{\mathrm{total}}$ is the total mass of stars formed in a grid of synthetic binary systems.
	\end{itemize}

In model set S1 the same physical assumptions are adopted as in the ``standard model'' of \cite{Claeys2014}.
The accretion efficiency of wind mass transfer is modelled according to the formulation of the BHL prescription
given by \cite{BoffinJorissen1988}. The wind is assumed to be spherically symmetric \cite[as in Eq. 4 of][]{Abate2013}
and consequently the relation between the angular momentum carried away by the expelled material, $\dot{J}$,
the total mass lost by the binary system, $\dot{M_1} + \dot{M_2}$,
and the orbital angular momentum of the system, $J_{\mathrm{orb}}$, can be written as
	\begin{eqnarray}
		\dot{J} = \gamma \times \frac{J_{\mathrm{orb}}}{M_1+M_2} \left(\dot{M_1} + \dot{M_2}\right)
		, \label{eq:jorbloss}
	\end{eqnarray}
%
where $\gamma=q=M_2/M_1$, that is the mass ratio of the accretor star over the donor star.

To investigate the effect of more efficient wind accretion from AGB donors, model set S2
calculates the wind accretion efficiency according to the WRLOF prescription proposed by \citet[][Eq. 9]{Abate2013}.

Our model set S3 uses the same accretion-efficiency model as set S2. In addition, the angular momentum lost 
by stellar winds is computed according to Eq. (\ref{eq:jorbloss}) with $\gamma$ defined as
	\begin{eqnarray}
		\gamma = \mathrm{max}\left\{~q~,~ h_{\rm BT93}~\right\}
		, \label{eq:bt93}
	\end{eqnarray}
%
where $h_{\rm BT93}$ has been determined by O.~R.~Pols (priv. comm.) from the best fit of the results of the 
ballistic simulations of \citet[][hereinafter BT93]{Brookshaw1993}. The factor $h_{\rm BT93}$ scales approximately 
as the ratio $(v_{\rm orb}/v_{\rm wind})^6$, where $v_{\rm orb}$ and $v_{\rm wind}$ are the orbital velocity of
the donor star and the wind velocity, respectively.
At wide separations $v_{\rm orb}\ll v_{\rm wind}$, the wind barely interacts with the orbit and the
spherically-symmetric-wind approximation applies. In contrast, for decreasing orbital separations the
ratio between $v_{\rm orb}$ and $v_{\rm wind}$ gradually increases and the angular momentum carried away
by the wind becomes progressively larger than in the isotropic-wind case. As a consequence, in model set S3
the binary systems typically widen less (compared to model sets S1 and S2), or even shrink, in response
to wind mass loss. Table~\ref{tab:models} presents a summary of the wind-accretion and angular-momentum
prescriptions adopted in each model set.

%

	\begin{table}
	\caption{\footnotesize{Wind-accretion efficiency and the angular-momentum prescriptions adopted in each model set.}}
	\label{tab:models}
	\begin{tabular}{ccc}
	\hline
	model set 	& wind accretion & angular momentum \\
	\hline
	S1 & BHL 	& isotropic wind\\
	S2 & WRLOF	& isotropic wind\\
	S3 & WRLOF	& based on BT93\\
	\hline
	\end{tabular}
	\end{table}

\section{Results}
\label{sec:results}

	\begin{figure}[t!]
		{\includegraphics[width=0.5\textwidth]{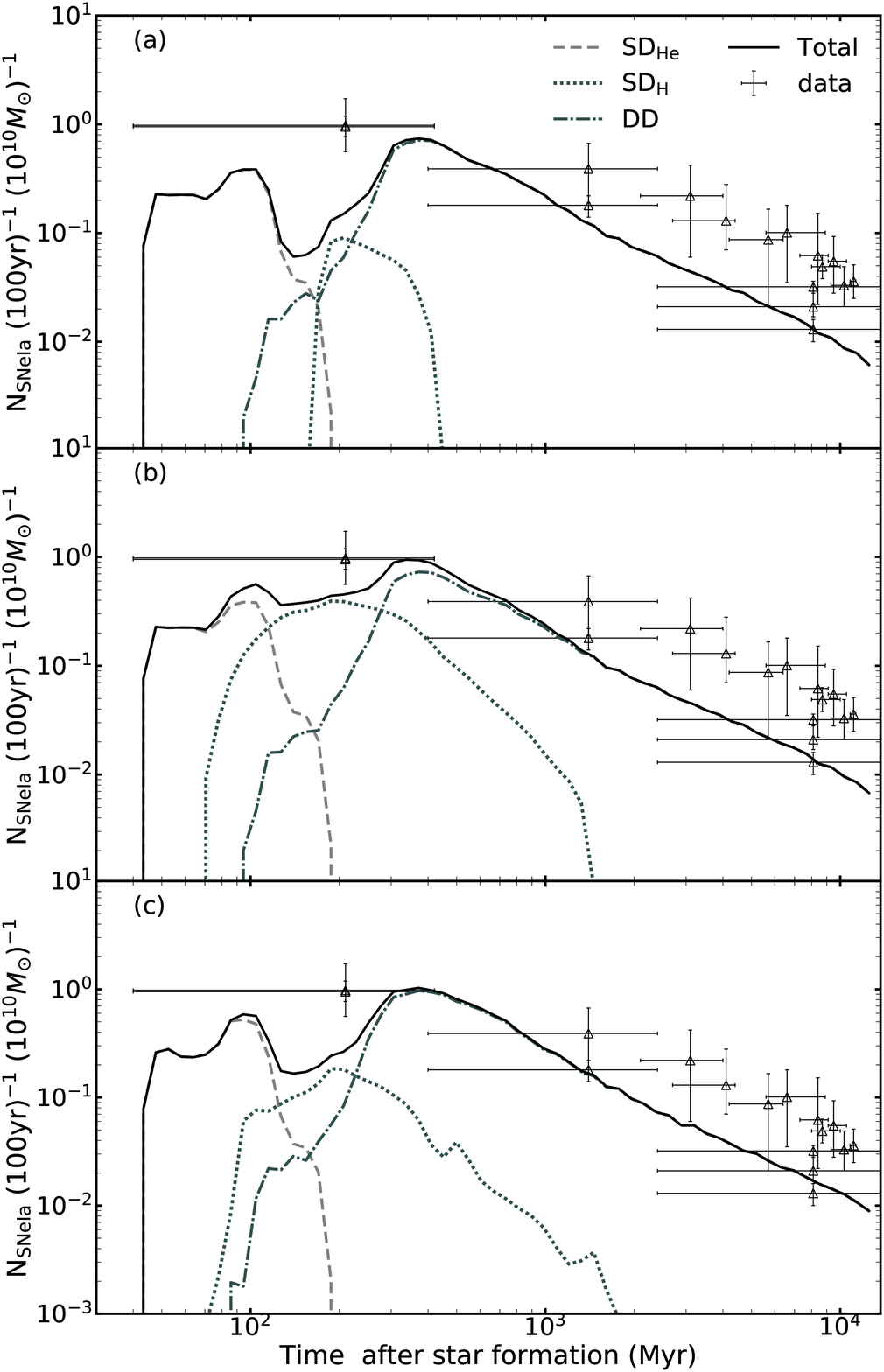}}
		\caption{\footnotesize 
				Delay-time distributions calculated with model sets S1, S2, and S3 (panels a, b, and c, respectively).
				The dashed and dotted lines represent the contributions to the \snia\ rate from the SD channel with He-rich 
				and H-rich donor stars, respectively, while the dot-dashed line shows the contribuion of the DD channel.
				The solid line shows the total predicted \snia\ rate.
				The open triangles are the literature data determined by \cite{Maoz2017}. 
				}
		\label{fig:DTD}
	\end{figure}

In Figs. \ref{fig:DTD}a--c the observed DTD \cite[open triangles, recently rederived by][]{Maoz2017} is compared 
with the predictions of model sets S1, S2 and S3, respectively. 
The individual contributions of different evolutionary paths, namely the SD channel with H-rich and He-rich donor stars 
and the DD channel, to the total \snia\ rate (solid line) are shown with the dotted, dashed, and dot-dashed lines, respectively. 
The \snia\ countings have been normalised per century and per $10^{10}\msun$ in stars.

Set S1 is the same as the default model of \citet{Claeys2014}, hence the features in Fig. \ref{fig:DTD}a are
essentially the same as in their Fig.~7. In particular, the contribution of He-rich donors (\SDHe) is prominent
at short delay times ($t<200$~Myr), whereas the DD channel dominates at delay times longer than about $300$~Myr.
The contribution of H-rich donors (\SDH) is marginal, as it is essentially restricted to stars on the first
giant branch transferring mass on the WDs by stable RLOF. 

In set S2, because of the larger range of orbital separations at which the AGB stars can efficiently
transfer mass onto the WD companions, their overall contribution to the \SDH\ channel becomes almost
ten times higher. The increase in wind-accretion efficiency has no influence on the \SDHe\ channel, 
in which mass transfer mostly occurs by RLOF, whereas it produces a small increase of the \snia\ rate
in the DD channel, because when the donor is an AGB star less mass is lost by the systems, hence the
orbital separations widen less and consequently a slightly higher proportion of binary systems end up 
with two WDs that are close enough to merge within a Hubble time.
In model set S2, the effect of AGB donors is mostly evident at delay times bewteen $100$ and $300$ Myr. 
At later delay times the DD channel dominates, as in set S1, although up to about $1$~Gyr a
small contribution to the total \snia\ rate comes from AGB donors within the \SDH\ channel.

The strong angular-momentum losses predicted by Eq. (\ref{eq:bt93}) cause model set S3 to form 
many more double WDs in close orbits than in sets S1 and S2. Consequently, the \snia\ rate in the
DD channel is about $35\%$ higher than in model S1. The contribution of the \SDH\ channel is increased
compared to model S1 but, in proportion, this channel is less important than in model S2. This occurs
because some of the systems, that in model S2 can efficiently transfer material from the AGB donor on
to the WD, shrink more in model S3 and consequently accrete less, because of the relation between
separation and accretion efficiency in the WRLOF model \cite[see e.g. Fig. 3 of][]{Abate2013}, and hence
the WD does not reach the Chandrasekhar mass.

\section{Discussion and Conclusions}
\label{sec:concl}

These results show that a less idealised treatment of wind mass transfer in binary systems significantly
increases the contribution of AGB stars in \snia\ progenitors. In particular, if a WRLOF prescription is
adopted for modelling wind accretion, AGB stars are the main donors in the SD channel and their
contribution is dominant between about $100$ and $300$ Myr after the initial burst of star formation.
In addition, a larger amount of angular momentum may be lost by the systems if the winds are not
ejected isotropically. For example, in model set S3 binary systems up to several thousands of days
shrink in response to wind mass loss, consequently more double WDs in close orbits are formed and the
contribution of AGB stars to the DD channel increases by a factor of three compared to the default case S1.

It should be emphasised that the simulations of BT93 only include gravitational effects on the wind 
particles, whereas the hydrodynamical properties of the gas are not taken into account. Consequently, the results
obtained with set S3 should be considered as a limit case of strong 
angular-momentum losses, which have sometimes been invoked to explain the orbital properties of the observed
carbon-enhanced metal-poor stars (\citealp{Abate2015-1} and \citealp[][in prep.]{Abate2017}).

In conclusion, these results suggest that AGB stars may play a significant role as progenitors of \snia, both
in the SD and DD formation scenarios, and detailed hydrodynamical simulations are required to reliably quantify
their impact on the total \snia\ rate.

\begin{acknowledgements} I thank O. R. Pols for sharing his angular-momentum model,
A.~Chiotellis for encouraging me to start this project, and the SOC\&LOC for a wonderful conference.
\end{acknowledgements}

\bibliographystyle{aa}

\end{document}